\def\today{\ifcase\month\or
  January\or February\or March\or April\or May\or June\or
  July\or August\or September\or October\or November\or December\fi

  \space\number\day, \number\year}

\magnification \magstep 1
\hsize 6 true in
\vsize 8.5 true in
\input amssym.def
\input amssym.tex
\catcode`\:=11
\newcount\bm:counta \newcount\bm:countb 
\newcount\bm:countc \newcount\bm:countd
\newtoks\bm:tok
\newif\ifbm:delim
\def\thehex#1{\ifcase\the#1 0\or 1\or 2\or 3\or 4\or 5\or 6\or 7\or
8\or 9\or A \or B\or C\or D\or E\or F\fi}%
\def\test#1#2{\ifcat#1#2\message{True}\else\message{False}\fi}
\newtoks \bm:savedtoks  \bm:savedtoks{}%
\def\bm:empty{\relax}%
\let\bm:save=\bm:empty
\newif\ifbm:cdr
\def\bm:split#1#2\bm:empty{%
    \def\bm:car{#1}\def\bm:cdr{#2\relax}%
    \expandafter\ifx\expandafter\relax\bm:cdr\bm:cdrfalse
    \else\bm:cdrtrue
    \fi}%
\newif\ifbm:found
\def\bm:in#1\find:#2\this:{%
    \def\find:##1#2##2##3\find:{%
        \ifx\bm:in##2\bm:foundfalse
        \else\bm:foundtrue
        \fi}%
    \find:#1#2\bm:in\find:}%
\let\when=\relax \let\use=\relax
%
%
\newif\ifboldwarning
\def\bm:message#1{{\newlinechar=`^^J
\immediate\write16{\string\bold\space warning on line
                                             \the\inputlineno^^J#1^^J}}}%
%
\def\latex:adjust{\expandafter\ifx\the\textfont0\csname twlrm\endcsname
                                           \def\bm:scale{1200}%
                              \else\expandafter\ifx
                                 \the\textfont0\csname elvrm\endcsname
                                                 \def\bm:scale{1095}%
                                               \else\def\bm:scale{1000}%
                                               \fi
                              \fi}%
\latex:adjust
\newdimen\bm:sevensize \newdimen\bm:fivesize
\bm:sevensize=.007pt \bm:fivesize=.005pt
\bm:sevensize=\bm:scale\bm:sevensize
\bm:fivesize=\bm:scale\bm:fivesize
\font\tenbf=cmbx10 scaled \bm:scale
\font\sevenbf=cmbx7 at \the\bm:sevensize
\font\fivebf=cmbx5 at \the\bm:fivesize
\textfont\bffam=\tenbf
\scriptfont\bffam=\sevenbf
\scriptscriptfont\bffam=\fivebf
:bit=cmbxti10 scaled \bm:scale
:bit=cmbxti10 at \the\bm:sevensize
:bit=cmbxti10 at \the\bm:fivesize
:scale
 at \the\bm:sevensize
 at \the\bm:fivesize
:bsf=cmssbx10 scaled \bm:scale
:bsf=cmssbx10 at \the\bm:sevensize
:bsf=cmssbx10 at \the\bm:fivesize
:bsl=cmbxsl10 scaled \bm:scale
:bsl=cmbxsl10 at \the\bm:sevensize
:bsl=cmbxsl10 at \the\bm:fivesize
:bmit=cmmib10 scaled \bm:scale
:bmit=cmmib7 at \the\bm:sevensize
:bmit=cmmib5 at \the\bm:fivesize
\newfam\bm:bmitfam
\textfont\bm:bmitfam=\tenbm:bmit
\scriptfont\bm:bmitfam=\sevenbm:bmit
\scriptscriptfont\bm:bmitfam=\fivebm:bmit
:bsy=cmbsy10 scaled \bm:scale
:bsy=cmbsy7 at \the\bm:sevensize
:bsy=cmbsy5 at \the\bm:fivesize
\newfam\bm:bsyfam
\textfont\bm:bsyfam=\tenbm:bsy
\scriptfont\bm:bsyfam=\sevenbm:bsy
\scriptscriptfont\bm:bsyfam=\fivebm:bsy
\newtoks\alphatok \alphatok{?}%
\expandafter\def\expandafter\new:fam\expandafter{\newfam}%
\def\set:fam#1{%
    \expandafter\ifx\csname #1fam\endcsname\relax
                    \expandafter\new:fam\csname #1fam\endcsname
                    \edef\alphafam{\csname #1fam\endcsname}%
                    \edef\set:fonts{%
                         \global\textfont\alphafam=\csname ten#1\endcsname
                         \global\scriptfont\alphafam=\csname
                                                seven#1\endcsname 
                         \global\scriptscriptfont\alphafam=\csname
                                                        five#1\endcsname}%
                    \set:fonts
                \fi
}%
\def\declare:alpha#1{%
    \set:fam{#1}%
    \expandafter\edef\csname math#1\endcsname{{%
                     \noexpand\if?\noexpand\the\noexpand\alphatok
                                  \global\noexpand\bm:savedtoks
                            {\noexpand\csname math:#1\noexpand\endcsname}%
                                  \noexpand\aftergroup\noexpand\bm:getarg
                     \noexpand\else\errmessage{You're already inside a
                                   \noexpand\expandafter\noexpand\string
                                   \noexpand\the\alphatok{..} - don't
                                   even think about it!}%
                     \noexpand\fi}}%
    \expandafter\def\csname math:#1\endcsname##1{{%
                     \alphatok\expandafter{\csname math#1\endcsname}%
                     \fam\csname #1fam\endcsname
                     \let\boldletter=\relax ##1}}}%
%
\chardef\rmfam=0
\chardef\mitfam=1
\chardef\calfam=2
\declare:alpha{rm}%
\declare:alpha{it}%
\declare:alpha{sl}%
\declare:alpha{tt}%
\declare:alpha{bf}%
\declare:alpha{mit}%
\declare:alpha{cal}%
\declare:alpha{sf}
\declare:alpha{bm:bmit}
\declare:alpha{bm:bsy}
\declare:alpha{bm:bsf}
\declare:alpha{bm:bit}
\def\default{\fam=-1 \def\boldletter##1{{\bf ##1}}}%
\let\mathbm=\mathbm:bmit :bsy
\let\mathbm:bcal=\mathbm:bsy
\def\bold#1{{
    \let\bm:currentsymbol=\relax
    \bm:split#1\bm:empty
    \ifbm:cdr\toks0{}\loop\toks0\expandafter\expandafter\expandafter{\expandafter\the\expandafter\toks0\expandafter\noexpand\expandafter\bold\expandafter{\bm:car}}\expandafter\bm:split\bm:cdr\bm:empty
                      \ifbm:cdr
                      \repeat
              \toks2\expandafter{\bm:car}%
              \xdef\bm:out{\the\toks0
                           \noexpand\bold\expandafter{\the\toks2}}%
              \aftergroup\bm:out 
    \else\ifx\bold#1\aftergroup\bold
         \else\ifmmode\bm:select{#1}%
                      \ifx\bm:save\bm:empty
                          \xdef\bm:out{\global\bm:savedtoks{}%
                             \the\bm:savedtoks\noexpand\bm:currentsymbol}%
                          \aftergroup\bm:out
                      \else\global\bm:savedtoks\expandafter\expandafter
          \expandafter{\expandafter\the\expandafter\bm:savedtoks\bm:save}%
                            \ifx\bm:currentsymbol\relax\aftergroup\bold
                            \else\let\test=F%
                                 \edef\argtest{\noexpand\bm:in
                                      \meaning\bm:currentsymbol
                                      \noexpand\find:
                                      \string\mathaccent
                                      \noexpand\this:}%
                                 \argtest
                                 \ifbm:found \let\test=T%
                                 \fi
                                 \edef\argtest{\noexpand\bm:in
                                      \meaning\bm:currentsymbol
                                      \noexpand\find:
                                      \string\radical
                                      \noexpand\this:}%
                                 \argtest
                                 \ifbm:found \let\test=T%
                                 \fi
                                 \if T\test\aftergroup\bm:getarg
                                 \else\aftergroup\bm:dumpchars
                                 \fi
                            \fi
                      \fi
              \else\errmessage{\string\bold\space should be used in
                               math mode only}%
              \fi
          \fi
    \fi
    }}%
\def\bm:getarg#1{{%
    \ifx#1\bold\aftergroup\bold
    \else\toks0{#1}\xdef\bm:out{\global\bm:savedtoks{}%
                        \the\bm:savedtoks{\the\toks0}}%
         \aftergroup\bm:out
    \fi}}%
\def\bm:dumpchars{{%
    \xdef\bm:out{\global\bm:savedtoks{}\the\bm:savedtoks}%
    \aftergroup\bm:out}}%
\def\bm:select#1{%
    \expandafter\bm:in\boldspecials\find:#1\this:
    \ifbm:found \def\when##1\use##2;{%
                    \ifx##1#1\xdef\bm:currentsymbol{\noexpand##2}%
                    \else\ifx##2#1\xdef\bm:currentsymbol{\noexpand##2}\fi
                    \fi}%
                \boldspecials
    \else\if\noexpand#1\relax                    
            \let\test=F%
            \edef\chartest{\noexpand\bm:in\meaning#1\noexpand\find:
                                         \string\mathchar\noexpand\this:}%
            \chartest
            \ifbm:found\let\test=T%
            \fi
            \edef\chartest{\noexpand\bm:in\meaning#1\noexpand\find:
                                           \string\char\noexpand\this:}%
            \chartest
            \ifbm:found \let\test=T%
            \fi                                  
            \if T\test 
          \expandafter\ifx\csname bold\string#1\endcsname\relax 
                          \edef\bm:process{\noexpand\defboldsymbol
                                           {\noexpand#1}%
                                           {\noexpand\mathchar}%
                                           {\the#1}{-1}}%
                          \bm:process
                       \fi \global\expandafter\let\expandafter
                           \bm:currentsymbol\expandafter=%
                           \csname bold\string#1\endcsname
            \else 
          \expandafter\ifx\csname bold\string#1\endcsname\relax 
                          \ifboldwarning\bm:message{Skipping \string#1.}%
                          \fi
                                        \def\bm:save{#1}%
                      \else\toks4\expandafter{%
                           \csname bold\string#1\endcsname}%
                           \edef\bm:save{\the\toks4}%
                           \global\expandafter\let\expandafter
                           \bm:currentsymbol\expandafter=%
                           \the\toks4
                      \fi
            \fi
         \else                                   
              \ifcat\noexpand#1A\gdef\bm:currentsymbol{\boldletter{#1}}%
              \else\ifcat\noexpand#1>            
             \expandafter\ifx\csname bold\string#1\endcsname\relax
                             \edef\bm:process{\noexpand\defboldsymbol
                                              {\noexpand#1}%
                                              {\ifnum\the\delcode`#1>-1
                                                 \noexpand\delimiter
                                               \else\noexpand\mathchar
                                               \fi}%
                                              {\the\mathcode`#1}%
                                              {\the\delcode`#1}}%
                             \bm:process         
                          \fi\global\expandafter\let
                             \expandafter\bm:currentsymbol
                             \expandafter=\csname 
                             bold\string#1\endcsname
                   \else\ifboldwarning\bm:message{Skipping \string#1.}%
                        \fi
                                      \def\bm:save{#1}%
                   \fi
              \fi
         \fi
    \fi}%
\def\defboldsymbol#1#2#3#4{%
    \bm:tok={#1}%
 \expandafter\ifx\csname bold\string#1\endcsname\relax
             \else\ifboldwarning\bm:message{Redefining
                                                \string\bold\string#1.}%
                  \fi
             \fi
    \bm:counta=#3
    \bm:countd=#4
    \ifnum\the\bm:counta="8000
           \expandafter\xdef\csname bold\string#1\endcsname{#1}%
    \else\ifx#2\delimiter \bm:delimtrue
         \else\ifx#2\radical \bm:delimtrue
              \else \bm:delimfalse
              \fi
         \fi
         \bm:countc=\bm:counta
         \divide\bm:countc by "1000             
         \advance\bm:counta by -\expandafter"\thehex\bm:countc 000
         \ifbm:delim\ifnum\the\bm:countd>-1     
                          \begingroup \bm:counta=\bm:countd
                          \divide\bm:counta by "1000
                              \begingroup
                              \multiply\bm:counta by "1000
                              \global\advance\bm:countd by
                                                      -\the\bm:counta
                              \endgroup
                          \bold:mathrecode
                          \multiply\bm:counta by "1000
                              \begingroup
                              \bm:counta=\bm:countd
                              \bold:mathrecode
                              \global\bm:countd=\bm:counta
                              \endgroup
                          \global\advance\bm:countd by \the\bm:counta
                          \endgroup
                      \else\ifboldwarning\bm:message{\the\bm:tok\space
                        is not a \string#2.^^JDoing the obvious thing..}%
                           \fi
                      \bold:mathrecode
                      \bm:countd=\bm:counta
                      \multiply\bm:counta by "1000
                      \advance\bm:countd by \the\bm:counta
                      \fi
                      \advance\bm:countd by \expandafter"\thehex
                                                        \bm:countc000000
                      \expandafter\xdef\csname bold\string#1\endcsname
                               {#2\the\bm:countd}%
          \else\bold:mathrecode
               \advance\bm:counta by \expandafter"\thehex\bm:countc 000
               \ifx#2\mathchar
                   \global\expandafter\mathchardef\csname
                                bold\string#1\endcsname=\the\bm:counta
               \else\expandafter\xdef\csname bold\string#1\endcsname
                               {#2\the\bm:counta}%
               \fi
          \fi
    \fi
}%
\def\bold:mathrecode{
      \bm:countb=\bm:counta
      \divide\bm:countb by "100
      \advance\bm:counta by -\expandafter"\thehex\bm:countb 00
      \ifcase\the\bm:countb
              \bm:countb=\the\bffam
      \or     \bm:countb=\the\bm:bmitfam
      \or     \bm:countb=\the\bm:bsyfam
      \else   \ifboldwarning\ifbm:delim\bm:message{Lack of bold
                                        extension fonts means
                                        \string\bold\the\bm:tok\space may
                                        not be bold.}%
                             \else\bm:message{Sorry, there just aren't the
                                   fonts for \string\bold\the\bm:tok.}%
                             \fi
              \fi 
      \fi
\advance\bm:counta by \expandafter"\thehex\bm:countb 00
                    }%
\def\DeclareBoldMacro#1#2#3{
    \bm:counta=#3 \bm:countd=\bm:counta
    \ifnum\the\bm:counta>"7FFF 
          \divide\bm:counta by "1000
    \else
          \multiply\bm:countd by "1000
    \fi
    \bm:countc=\bm:counta
    \divide\bm:countc by "1000
    \advance\bm:countd by -"\thehex\bm:countc 000000
    \edef\bm:process{\noexpand\defboldsymbol{\noexpand#1}{\noexpand#2}%
                     {\the\bm:counta}{\the\bm:countd}}%
    \bm:process}%
%
%
\DeclareBoldMacro{\{}{\delimiter}{"4266308}
\DeclareBoldMacro{\}}{\delimiter}{"5267309}
\DeclareBoldMacro{\langle}{\delimiter}{"426830A}
\DeclareBoldMacro{\rangle}{\delimiter}{"526930B}
\edef\FixLessThanGreaterThan{\noexpand\DeclareBoldMacro{<}%
{\noexpand\mathchar}{\the\mathcode`\<}%
\noexpand\DeclareBoldMacro{>}{\noexpand\mathchar}{\the\mathcode`\>}}%
\FixLessThanGreaterThan
\DeclareBoldMacro{\sqrt}{\radical}{"270370}
\default
\def\boldspecials{\when\mathmit\use\mathbm:bmit;\when\mathcal\use\mathbm:bcal;\when\mathrm\use\mathbf;\when\mathsf\use\mathbm:bsf;\when\mathit\use\mathbm:bit;\when\mathtt\use\mathtt;\when\mathsl\use\mathbm:bsl;\when\default\use\default;}%
\boldwarningtrue
\catcode`\:=12
\overfullrule=0pt
\def\beginsection#1\par{\bigbreak\leftline{\bf #1}\nobreak\smallskip\noindent}
\openup 1 \jot
\centerline {\bf THE SEN  CONJECTURE  FOR}
\centerline {\bf  FUNDAMENTAL
MONOPOLES OF DISTINCT TYPES}
\vskip 1.5 cm
\centerline {G.W. GIBBONS}
\centerline {D.A.M.T.P.}
\centerline {University of Cambridge}
\centerline {Silver Street}
\centerline {Cambridge CB3 9EW}
\centerline {U.K.}
\vskip 1.5cm

\centerline  {\bf ABSTRACT}
\tenrm
{\narrower \narrower \smallskip
I exhibit a middle-dimensional square integrable  
harmonic form on  the moduli space of  distinct fundamental BPS  monopoles of 
an arbitrary Lie group. This is in accord with Sen's S-duality conjecture.
I also show that the moduli space has no closed or bound geodesics }

\beginsection I. Introduction

There have recently been some advances in our understanding
of the interactions of BPS monopoles  in the case
that a compact semi-simple Lie group $G$ of rank $k$
breaks
to its maximal torus $C(G) \cong U(1)^k$ by a Higgs field $\bold \Phi$
in the adjoint  representation [1,2,3,4]. 
In the  simplest non-trivial case ($G=SU(3)$)
the metric on  the relative moduli space has been identified as the Taub-NUT
metric and a square integrable self-dual
harmonic form
exhibited which is consistent with Sen's S-Duality
conjecture [5]. A proposal has been made [4] for the metric
on the general moduli space of distinct fundamental monopoles
but the corresponding middle-dimensional harmonic form
was not found. In this paper I shall remedy that deficiency by
giving a simple explicit expression for this form
which represents a bound state of a system of fermions and monopoles
in the $n=4$ supersymmetric version of the theory.
I shall also show that in distinction
to the case of arbitrary numbers of identical $SU(2)$ monopoles, there
are no classical closed geodesics. Indeed
there are no classical bound orbits at all.   

\beginsection Fundamental Monopoles

Monopoles in this theory may carry $k$ types of magnetic charge.
One may always arrange, by means of a conjugation if necessary,
that the vacuum expectation value of the Higgs field 
at infinity ${\bf \Phi}_\infty$
lies in the Cartan sub-algebra $\frak {h}$.
Associated with ${\bf \Phi}_\infty$ is a hyperplane in $\frak{h}$ and
a unique set of $k$ simple positive  roots ${\bold \beta}_ a$, $a=1,2\dots,k$.
A general monopole has an associated magnetic charge vector ${\bf g}$
taking values in ${\frak{h} }^\ast$, the dual of the Cartan sub-algebra. 
In fact the Dirac quantization condition dictates that $\bf g$
lies on a lattice spanned by the reciprocal vectors  
$$ {\bold \beta}_ a^\ast = { {\bold \beta}_ a \over {\bold \beta}_ a.{\bold \beta}_ a,}
\eqno(1)
$$
of the positive simple roots, that is
$$
{\bf g} = { 4 \pi \over e} \sum ^{a=k}_{a=1}n_a {\bold \beta}_ a^\ast,
\eqno(2)$$
where $e$ is the gauge coupling constant and $\{n_a\}$ are integers.

There are  $k$ types of  {\it fundamental}
monopoles corresponding to an embedding of
an $SU(2)$ monopole  and each has  unit magnetic charge with respect to
one of the $k$ circle subgroups of $C(G)$ and zero magnetic charge  with
respect to the other $k-1$ circle subgroups. The $a$'th fundamental
 monopole is associated with the dual root ${\bold \beta}_ a^\ast$ 
and has positive mass 
$$
m_a= { 4\pi \over e}  {\bold \beta}_ a^\ast . {\bf \Phi}_\infty .
\eqno(3)$$

Remarkably  a sort of superposition of distinct fundamental monopoles
is possible. These correspond to a composite dual root
$$
{\bold \alpha }^\ast = \sum ^{a=k}_{a=1} n_a {\bold \beta}_ a^\ast
\eqno(4)$$
and have mass
$$
m= \sum ^{a=k}_{a=1} n_a m_a ,
\eqno(5)$$
where the integers $n_a$ are non-negative.

We shall consider the moduli space of $n$ distinct fundamental
monopoles corresponding to $n$ positive roots ${\bold \alpha}_i$.
The $n$ positive roots define a Dynkin diagram.
The asymptotic forces between the monopoles  are given 
entirely in 
terms of the 
inner products ${\bold \alpha}_i.{\bold \alpha}_j \ne 0$
Only roots which are connected in the Dynkin diagram (i.e. for which
${\bold \alpha}_i.{\bold \alpha}_j \ne 0$) interact and thus 
one is led to restrict attention to 
 Dynkin diagrams which are  connected and therefore the ${\bold \alpha}_i$
constitute the roots of (possibly smaller) group. The Dynkin diagram has
just $n-1$ links correponding to the number of unordered pairs $(i,j)$
for which  ${\bold \alpha}_i.{\bold \alpha}_j \ne 0$.
We shall use the index $A$ to label these links.
Physically the links correspond  to $n-1$ relative position vectors
${\bf r} _A = {\bf x}_i-{\bf x}_j$ and $n-1$ relative phases 
$\psi_A \in (0,4\pi ]$.  
For each link we define the positive numbers 
$$
\lambda _A= -2 {\bold \alpha}_i^\ast .{\bold \alpha}_j ^\ast. 
\eqno(6)$$
\beginsection The moduli space.    

In general the moduli space ${\cal M}_n$ of $n$ BPS monoples
is known to be a 
 $4n$-dimensional geodesically complete HyperK\"ahler manifold. 
Because the centre of mass motion may be factored out, it is  of the form 
$$
{\cal M} _n \cong {\Bbb R}^3   \times { S^1 \times {\cal M} ^{\rm rel}_{n-1} \over {\cal D} }
\eqno(7)$$
where the relative moduli space $ {\cal M} ^{\rm rel}_{n-1} $ is
a geodesically complete $4(n-1)$-dimensional hyperK\"ahler manifold
and the group $\cal D$ is a 
discrete normal subgroup of the isometry group of 
${\cal M} ^{\rm rel}_{n-1}$. 
The  isometry group contains a copy of   $SO(3)$  acting on the 
the three complex structures as a triplet.
In general  ${\cal M} ^{\rm rel}_{n-1}$ will have 
 no additional exact continuous isometries  but
for large separation one may
identify as coordinates 
$n-1$  relative cartesian positions ${\bf r}_A$ and $n-1$ and certain angles 
$\theta_A $ as coordinates on ${\cal M}^{\rm rel }_{n-1}$. 
One then finds that asymptotically there is an additional 
approximate triholomorphic action
of the torus group $T^{n-1}$ 
corresponding to shifting the angles. The associated Killing vector fields
are
$$
K^{A \alpha} {\partial \over \partial x^ \alpha } 
= {\partial \over \partial \theta _A} 
\eqno(8)$$
with $\alpha = 1,2\dots 4n-4$. The invariance corresponds 
physically to the conservation of
the $n-1$ relative electric charges which may be carried by
dyons. 
One may explicitly write down the asymptotic
metric $g_{\alpha \beta}$ in terms of magnetic charges of the monopoles [6,7].
In the rest of this section
I will describe some general properties 
of the asymptotic metric. In the case of distinct monopoles
the asymptotic metric is believed to be exact
and so these properties are  shared by the exact metric.
Thus for example there  is always an asymptotic identification
between  the space of relative positions
and  the quotient of $ {\cal M}^{\rm red} \equiv
M^{\rm rel}_{n-1} /T^{n-1}$. This identification is believed to be exact
in the case of distinct monopoles.
The cartesian coordinates ${\bf r}_A$ are, up to a scale, the three
moment maps corresponding to the triholomorphic Killing fields $ \partial
\over \partial \theta_A$. The torus action on ${\cal M}^{\rm rel}_{n-1}$is not even locally
hypersurface orthogonal and so regarding $ 
{\cal M} ^{\rm rel}_{n-1} $ as a $T^{n-1}$ bundle over $ {\cal M}^{\rm red} $
one gets a non-trivial torus connection  in the standard
Kaluza-Klein fashion by taking the horizontal subspaces to 
be orthogonal to the torus fibres. 
Thus  a curve $x^\alpha (t)$ in $M^{\rm rel}_{n-1}$ is horizontal
if 
$$
 K^{ A \alpha} g_{\alpha \beta }  { dx^\beta \over dt} =0.
\eqno(9)$$
 
The curvature 2-forms  
$F_A$ on $M^{\rm rel}_{n-1}$
are given in terms of the Killing co-vector fields
$K ^A _\alpha = g_{\alpha \beta} K^{A \beta}$ by
$$
F^A = \partial _ \alpha K^A_\beta -\partial _ \beta K^A_\alpha.
\eqno(10)$$
It is a well known elementary deduction from Killing's equations
that if a  metric $g_{\alpha \beta}$ is Ricci flat then a two-form
obtained by taking the exterior derivative of
the one-form obtained by lowering the index
of any Killing vector field
is both  closed and co-closed, that is it satisfies Maxwell's 
equations. Since a HyperK\"ahler metric is necessarily
Ricci flat it follows that we obtain $n-1$ solutions of Maxwell's equations
in this way. They will be important later.  

Any $SO(3) \times T^{n-1}$ invariant metric, whether HyperK\"ahler
or not, may locally be cast in the form
$$
ds^2 = G_{AB} d{\bf r}_A . d {\bf r}_B + 
H^{AB} ( d \theta _A + {\bf W}  _{AC}.  d {\bf r}_C )( d \theta _B + {\bf W} _{BD}. d {\bf r}_D ),
\eqno(11)$$
where $G_{AB} $, $H^{AB}$  and ${\bf  W }_{AC}$ are independent of the
angles $\theta _A$.  
The Maxwell vector potentials are then given by
$$
K^A = H^{AB}  ( d \theta _B + {\bf W} _{BD } . d {\bf r}_D ).
\eqno(12)$$

In the particular case that the metric is HyperK\"ahler
one has (among other things) that 
$$
G_{AB} = H_{AB}
\eqno(13)$$
where $H_{AB} H^{BC}= \delta _A^C$.
\beginsection Taub-NUT Space

It may be helpful to begin by considering the simplest example when $G=SU(3)$
and  $n=2$.
In this case  $M^{\rm rel}_1 $ has been identified [1,2,3] as the complete
self-dual
Taub-NUT space on ${\Bbb R}^4 \equiv {\Bbb H}$,
with the isometry group $U(2)$ acting  on ${\Bbb H}$ by left
 multiplication by a unit quaternion and right multiplication
by a circle subgroup of the quaternions.  Each of the two-sphere's
worth of  
 complex structures act by left multiplication by a unit quaternion
and the circle action commutes with this action, i.e. it is
 triholomorphic.  The $SO(3)$ action however rotates the complex 
structure.  If $\sigma_1, \sigma_2 , \sigma_3$ are 
left-invariant one-forms on $SU(2)$ the metric takes the form
$$
ds^2 = V^{-1} 4 M^2 \sigma_3^2 + V \Bigl ( dr^2 + r^2 (\sigma _1^2 + \sigma _2 ^2) \Bigr ) 
\eqno(14)$$
$$
 = V^{-1} (d \tau + {\bold \omega}. d{\bf r})^2 + V d{\bf r}.d{\bf r}
\eqno(15)
$$
 with $V= 1+{2 M \over r} $, $\nabla \times{\bold  \omega}  = \nabla V$
and $d\tau + {\bold \omega}. d{\bf r}= 2M \sigma _3 = 2M ( d \psi + \cos \theta d \phi )$, where $(\psi , \theta , \phi)$ are Euler angles and $M$ is
a positive constant.
The metric has a coordinate singularity at $r=0$ because the isometry
group has a fixed point at the origin, sometimes referred to
as a $NUT$. As stated above the three cartesian coordinate functions  $\bf r$ 
which parameterize the orbits of the triholomorphic circle action may be 
invariantly characterized (possibly up to an overall scale)
as its  three moment maps.

Because it is the exact metric
the  manifold must be complete and therefore
the mass parameter $M$ which appears in the Taub-NUT metric must,
as stated above  be {\sl positive}
in distinction to the asymptotic form of the relative moduli space
of  two identical $SU(2)$ monopoles for which the mass parameter is 
negative. This change of sign reflects the different nature of the interactions between monopoles of the same type and those of different types.
In general the former are attractive and the latter are repulsive.
As we shall see later this leads to significant differences in the 
behaviour of geodesics and bound states.

It is worth remarking here that although  the negative mass Taub-NUT metric
is not complete it is not really pathological
and has as good a geometrical pedigree as the positive mass
metric.
The problem is that  the signature 
of the metric changes from $+ + + +$ to $- - - - $ as one passes
$r=2|M|$. The region inside $r=2|M|$ is complete near $r=0$,
which is just a coordinate singularity. To understand the change of 
signature at $r=2|M|$ recall that  
the Taub-NUT metric with positive mass may be obtained as a hyperK\"ahler
 quotient of the flat HyperK\"ahler metric  on
${\Bbb R}^8$. The zero set of the relevant moment map is a smooth
5-dimensional submanifold  in ${\Bbb R}^8$. One then quotients
by the  ${\Bbb R}$-action generated by the moment maps.
 To get the Taub-NUT metric with negative  mass one starts
with  the flat pseudo-HyperK\"ahler metric  on
${\Bbb R}^{4,4} $. The zero set of the relevant moment map
is again a smooth
5-dimensional submanifold  but now in ${\Bbb R}^{4,4}$. The  metric induced
 on it
from the pseudo-euclidean metric of signature 
$++++----$  has signature $++++-$ if $r>2|M|$
and signature $----+$ if $r<2 |M|$. 
The orbits of the ${\Bbb R}$-action generated by the moment maps
are timelike if $r> 2 |M|$ and spacelike if $r<2 |M|$. As a consequence
the quotient has signature $++++$ if $r>2 |M|$ and $----$ if
$r<2|M|$. Of course the induced metric vanishes on $r=2|M|$.

To return to the positive mass case: an $SO(3)$-invariant  square integrable 
2-form $F$ exists in Taub-NUT which is both closed and co-closed [2,3].
In fact the two-form found in [2,3] is {\sl precisely}  that obtained
from the $U(1)$ Killing field. It is exact,
$$
F=dA,
\eqno(16)$$
and  the globally well-defined
one-form $A_\alpha$ is related to the Killing vector field
$K^\alpha {\partial \over \partial x^ \alpha} $ which generates the tri-holomorhic circle action
by index lowering with the metric $g_{\alpha \beta}$
$$
A_\alpha = g_{\alpha \beta} K^{\alpha}.
\eqno(17)$$
Explicity  
$$
A= 4M^2 V ^{-1} \sigma _3.
\eqno(18) $$
Because $V \rightarrow 1$ at large $r$,
the length of the orbits of the circle action,
which is proportional to the norm of $A_\alpha$,  tends to
constant  at infinity. Therefore  the one form $A$ is not 
square integrable.

\beginsection The General Moduli-space Metric

Lee, Weinberg and Yi [4] have proposed 
in the case of $n \le k $ fundamental monopoles 
for any group $G$  
that the asymptotic moduli space metric at large separations
is in fact exact for all separations. As explained above
there is no loss of generality in taking $n=k$.
The metric is specified by
$$
G_{AB} = \mu_{AB} + {g^2 \lambda _A  \delta _{AB} \over
8 \pi r_A}  
\eqno(19)$$
where $r_A=|{\bf r}_A|$ and $g= {4 \pi \over e}$ and
 $\mu_{AB} $ is 
a constant positive definite non-diagonal symmetric matrix. 
The angles $\theta _A$ are given in terms of the phases 
$\psi _A \in (0,4\pi] $  by 
$$
\theta_A = {g^2 \lambda _A \over 8 \pi} \psi _A
\eqno(20)$$
where the  $\lambda_A$ are given by (6) and 
$$
{\bf  W }_{AC}=  {g^2 \lambda _A \delta _{AC} \over 8 \pi} {\bf w} ({\bf r}_A),
\eqno(21)$$
where ${\bf w} ({\bf r})$ is the vector potential of a single Dirac 
monopole. Note that the repeated indices in (21) are not summed over
and  note also that there is a single radial
magnetic  field ${\bf B}_A$ 
associated  with each link.

Lee, Weinberg and Yi have checked that the apparent singularities
arising when one or more relative distances $r_A$ vanishes
is a coordinate artefact and that the metric is in fact complete there.
This is certainly a necessary test of their conjecture.
They did not discuss the global topology.  
In the case $G=SU(k+1)$ it is known from other arguments [1]  
 that topologically ${\cal M}^{\rm rel}_{k-1}  \cong {\Bbb R}^{4k-4}$
and it seems likely that this is true in general. This agrees
with a count of the fixed points of the $U(1)$  actions.
One might worry that the proposed metric is just the metric product
of $k-1$ copies of the Taub-NUT metric but this appears 
not to be the case because the matrix $\mu_{AB}$ and hence 
the matrix 
$G_{AB}$ is not diagonal.
 
\beginsection Harmonic Forms

The square integrable harmonic two form on the Taub-NUT
metric has an obvious analogue on the Lee-Weinberg-Yi metric.
As pointed out above each of the $k-1$ $U(1)$  Killing vector fields
$K^{A\alpha}$  provides a smooth, exact and co-closed $SO(3)$
-invariant two form $F^A$ . 
This dies away like $1 \over r_A^2$ at infinity. 
The volume is easily seen to grow as  
$$
\prod ^{A=k-1}_ {A=1} r^3_A
\eqno(22)$$
at infinity and therefore $F^A$  is not square integrable.
We can obtain higher rank closed and co-closed even-dimensional
forms by taking sums  of exterior  products but it is clear that there
is only one way of constructing a square integrable form in this way
: one must take the middle-dimensional form:
$$
F= \prod ^{A=k-1}_ {A=1}  F^1 \wedge F^2 \dots .
\eqno(23)$$
It is easy to see that although the $2(k-1)$ -form $F$ is exact, 
the $(2k-3)$-form of which it is the exterior derivative is not
square integrable. Finally it is clear that this middle dimensional
form, which is manifestly $SO(3)\times T^{k-1}$ invariant,
 is self-dual, in other words
$$
F=\star F 
\eqno(24)$$
where $\star$ is the Hodge star operation on forms. 
Thus  the $2(k-1)-$ form $F$ satisfies 
all the properties  predicted 
by S-duality whose construction was
left as an outstanding problem by Lee, Weinberg and Yi. 

To be strictly accurate there is an issue of uniqueness.
In fact there is, as far as I know,  no 
 rigorous proof
that the square integrable harmonic form on the relative 
moduli space of two identical $SU(2)$ monopoles
is unique. The arguments given in [5] assume $SO(3)$ invariance
and while it is clear by averaging that if a harmonic 
form exists
there also exists an $SO(3)$ invariant one it is not
obvious without a further argument that every harmonic form
is $SO(3)$ invariant. A similar statement  is true in the present case
with respect to $SO(3)\times T^{k-1}$ invariance.

To check uniqueness among non-$SO(3)\times T^{k-1}$-invariant 
forms one might procede
by  considering  the difference 
between $F$ 
and a putative rival $G$ say. One has
$$
F-G=dB
\eqno(25)$$
for some  globally defined $2k-3$-form $B$ which satisfies
$$
\delta d B=0.
\eqno(26)$$
Contracting with $B$ and integrating over the part of ${\cal M}^{\rm rel}_{k-1}$ inside
a large bounding hypersurface $\partial {\cal M}^{\rm rel}_{k-1}$ gives
$$
\int_{{\cal M}^{\rm rel} _{k-1}} || dB ||^2 = \int _{\partial {\cal M}^{\rm rel} _{k-1} } B \delta B . 
\eqno(27)$$
The area of the boundary $\partial {\cal M} ^{\rm rel} _{k-1} $ increases 
as
$$
\prod ^{A=k-1}_{A=1} r^2 _A.
\eqno(28)$$
Evidently if $B$ and $\delta B$ decrease  faster
than
$$
\prod ^{A=k-1}_{A=1} r^{-1} _A,
\eqno(29)$$ 
then we can deduce uniqueness: $dB=F-G=0$. It is not difficult
to convince one's self that these fall-off
conditions are very plausible  but I have no 
truly rigorous  proof.

\beginsection Absence of Bound Orbits and the Virial Theorem

The original reason for being interested
in the metric on the moduli space is that the geodesics
give an approximate description of the slow motion
of monopoles. At large separations this is in fact already known from
a consideration of the forces betwen widely separated monopoles. 
Indeed that is how the asymptotic metric is constructed!
In the case of fundamental monopoles the asymptotic metric is exact
and thus use of the  metric could be avoided if one wishes.
The motion of the monopoles remains of interest however. 
From a physical point of view it is simplest
to consider the projections of geodesics
onto the  
quotient space ${\cal M}^{\rm red}$ and to ignore the 
motion in the torus fibres. The effect of the latter
is to endow the magnetic monopoles with conserved 
electric charges 
$$
Q^A= K^A _\alpha { dx ^ \alpha \over dt} = H^{AB} (\dot {\theta} _B+
{\bf W}_{BC} .{\bf v}  _C ), 
\eqno(30)$$ 
where $ {\bf v}  _C= {d  { \bf x}  _C \over dt}$
is the velocity in the reduced space ${\cal M}^{\rm red}$.  
Given an orbit  in the reduced space we can reconstruct the motion
in the angles $\theta_A$.  

In the case of Taub-NUT the reduced motion has a very simple description [8].
Angular momentum conservation implies that the orbits lie on a cone centred
on the origin. The existence of a generalization of the  
conserved Lagrange-Laplace-Runge-Lenz vector for the Coulomb problem
 then implies that the orbits lie in a plane. 
As a consequence they are conic sections. If the mass parameter is positive,
 as it is for two distinct
fundamental $SU(3)$ monopoles, there are only hyperbolic orbits.
 If the mass parameter is negative as it is for the
approximate metric for two  identical $SU(2)$ monopoles
then bound elliptical orbits are possible. The coresponding
 bound geodesics 
persist on the exact metric which is also  known 
to support closed geodesics [9]. Recently
Bielawski [10] has given a totally geodesic
embedding of the strongly centred $SU(2)$ monopole moduli space
of charge 2  into the 
strongly centred$ SU(2)$  moduli space of charge $n $ thereby showing the existence
of closed geodesics for all $n$.   

The reason for this difference  is essentially
because $SU(2)$  dyons are oppositely charged with 
respect to the same $U(1)$
which gives rise to attractive forces between the dyons.
In the case of distinct fundamental $SU(3)$ monopoles the dyons are 
charged with respect to different $U(1)$'s and in fact
the forces are repulsive.   

It is clearly of interest to ask whether there are bound orbits,
and hence bound geodesics,  for more than
two distinct fundamental monopoles. In fact there are not.
To prove this we establish a simple generalization of the
Virial Theorem. We can obtain our result
directly from the geodesic equations of motion
but some extra insight is afforded by constructing
an  effective Lagrangian $L$ from which to obtain the equations of motion of the orbits
with fixed electric charges
 on ${\cal M}^{\rm red}$. The Lagrangian (which is {\sl not} just
obtained by substituting the expressions for conserved charges
into the action for geodesics)  is readily 
 seen to be given by
$$
L= { 1\over 2} G_{AB} {\bf v}_A .{\bf v}_B - { 1\over 2} H_{AB} Q^A Q ^ B
+ Q^A {\bf W}_{AB} . {\bf v}_B. 
\eqno(31)$$
The conserved energy is 
$$
E = { 1\over 2} G_{AB} {\bf v}_A .{\bf v}_B +{ 1\over 2} H_{AB} Q^A Q ^ B.
\eqno (32)$$
The two terms in(32) correspond to the kinetic energy  
$$
T= { 1\over 2} G_{AB} {\bf v}_A .{\bf v}_B ,    
\eqno(33)$$
which is positive definite and the potential energy
$$
V= { 1\over 2} H_{AB} Q^A Q ^ B,
\eqno(34)$$
which is also positive and repulsive for distinct fundamental monopoles.
It follows immediately that the relative distance $r_A$ can only vanish
if the associated electric charge $Q^A$ vanishes.
Because magnetic fields do no work, the vector potentals  
$ {\bf W} _ B = Q^A {\bf W}_{AB} $ are absent from the expression for the energy. 
The equations of motion are
$$
- { d \over dt } \bigl ( G_{AB}{\bf v}_B \bigr ) + 
{\partial T \over \partial {\bf r} _A}
- {\partial V \over \partial {\bf r} _A} - {\bf B} _A \times {\bf v}_A=0.
\eqno(35)$$
Note that although the index $B$ is summed over, there is no sum over the index $A$. One now takes the dot product with ${\bf r}_A $.
Because of the special form of the vector potential we are considering 
the last term vanishes. Thus after a slight re-arrangment
(again with a sum over $B$ but not $A$)
$$
{ d \over dt } \bigl ( G_{AB}{\bf v}_B .{\bf r}_A \bigr ) =  
G_{AB}{\bf v}_B .{\bf v}_A + 
{\bf r}_A. {\partial T \over \partial {\bf r} _A}- 
{\bf r}_A.  {\partial V \over \partial {\bf r} _A}.
\eqno(36)$$

At this stage (although one loses some information)
 it is quickest to sum over $A$ and use the homogeneity
properties with respect to the ${\bf r}_A$ of the 
kinetic and potential energies.
One has
$$
T= T_0 + T _{-1}
\eqno(37)$$
and 
$$
V=V_0 + V_{-1}
\eqno(38)$$
where $T_0= { 1\over 2} \mu _{AB} {\bf v}_A. {\bf v} _B$
 and $V_0= { 1\over 2} \mu _{AB} Q^A Q^B$  are postive and
independent
of position ($V_0$ is just a constant) and $T_{-1}$ is homogeneous of
degree $-1$ in the ${\bf r}_A$'s and $V_{-1}$ is also positive and homgeneous
of degree $-1$ in the ${\bf r}_A$'s.
Thus 
$$
{ d \over dt} \bigl ( G_{AB}{\bf v}_B .{\bf r}_A \bigr ) = 2T_0 +T_1 + V_{-1} = E^\prime + T_0, 
\eqno(39)$$
where 
$$
E^\prime = E-V_{0}= T+ V_{-1}
\eqno(40)$$
is a strictly positive constant. One now integrates
(39)  from an initial time $t_i$  to a final time $t_f=
t_i+P$ and divides by $P$. If one has a bound orbit
then the  right hand side
must go to zero  for large enough $P$ but the right hand side is never smaller than 
the positive constant $E^\prime$. This is a contradiction. 
This argument also rules out periodic orbits but in that case,
of course, one could choose $P$ to be the period of the orbit.
If the forces were attractive and if bound orbits
existed  we  would obtain in this way a relation between the 
average kinetic
and potential
energies. This is a generalization of the  usual Virial Theorem.

The non-existence of bound classical motions indicates that 
there are no purely bosonic quantum bound states, in distinction to
 the $SU(2)$ case where it is  known that there are non-BPS bound states [7]. Thus the existence of the Sen bound state
in the case of distinct fundamental  monopoles is a more unexpected
phenomenon  and  appears to
owe its existence to a 
more subtle consequence
of the fermionic structure of the theory.

\beginsection   Acknowledgement

I would like to thank Nick Manton
for helpful conversatiions and a critical reading of the manuscript.

\beginsection References

\medskip \item {[1]} S A Connell "The Dynamics of the $SU(3)$ Charge (1,1) Magnetic Monopole" University of South Australia Preprint 
\medskip \item {[2]} J P Gauntlett and D A Lowe "Dyons and S-Duality
in $N=$ Supersymmetric Gauge Theory" hep-th/960185 
\medskip \item {[3]} K Lee, E J Weinberg and P Yi "Electromagnetic Duality
and $SU(3)$ Monopoles" hep-th/9601097
\medskip \item {[4]} K Lee, E J Weinberg and P Yi 
"The Moduli Space of Many BPS Monoples for Arbitrary Gauge Groups" hep-th/9602167 
\medskip \item {[5]} A Sen  {\sl Phys Lett} {\bf B 329 }(1994) 217-221

\medskip \item {[6]} N S Manton {\sl Phys Lett} {\bf B 154 }(1985) 397-400
: {\bf B 157 }(1985) 475 (E)
\medskip \item {[7]} G W Gibbons and N S Manton {\sl Phys Lett} {\bf B 356}
(1995) 32-38 
\medskip \item {[8]} G W Gibbons and N S Manton {\sl Nucl
Phys } {\bf B 274} (1986) 183-224 
\medskip \item {[9]} L Bates and R Montgomery {\sl Comm Math Phys} {\bf 118}
(1988) 635-640   
\medskip \item {[10]} R Bielawski "Existence of Closed Geodesics on the 
Moduli Space of $k$-Monopoles" McMaster University Preprint

\bye